\documentstyle[aps,epsfig]{revtex}
\begin{document}
\twocolumn[\hsize\textwidth\columnwidth\hsize\csname
@twocolumnfalse\endcsname

    \title{\Large\bf GPS observables in general relativity}
    \author{Carlo Rovelli\\[.2cm] {\it Centre de Physique 
    Th\'eorique de  
    Luminy, F-13288 Marseille, EU}\\
     {\it Department of Physics, Pittsburgh University, Pittsburgh  
     PA-15250 USA}}\date{\today}  \maketitle
 
\begin{abstract} 
I present a complete set of gauge invariant observables, in the
context of general relativity coupled with a minimal amount of
realistic matter (four particles).  These observables have a
straightforward and realistic physical interpretation.  In fact, the
technology to measure them is realized by the Global Positioning
System: they are defined by the physical reference system determined
by GPS readings.  The components of the metric tensor in this physical
reference system are gauge invariant quantities and, remarkably, their
evolution equations are local.  \end{abstract} \vskip1cm]

In general relativity (GR), the ``observables", namely the physical
quantities that we can predict and measure in real experiments,
correspond to quantities of the theory that are invariant under
coordinate transformations.  It is not difficult to construct
observables in concrete applications of the theory.  Indeed, each
physically meaningful number that we actually measure corresponds to
one of these observables.  For instance, in studying the general
relativistic dynamics of the solar system, the distance of a planet
from us, at given (solar) time (say, at midnight tonight) is an
observable.  On the other hand, most of the theoretical work is not
done in terms of observables, but rather in terms of gauge dependent,
that is, coordinate dependent, quantities.  This of course is the
reason for which the same physical situation can be described in terms
of different metrics tensors.  The usual procedure in GR is indeed to
develop the theoretical description of a certain physical situation in
an arbitrary gauge, and then compute the value of a certain number of
coordinate independent observables, which can be compared with 
observations.

This way of proceeding has an unsatisfactory aspect: we do not have at
our disposal a {\em complete\/} set of observables, which we could
imagine, in principle, to measure in a simple manner.  A physical
situation, of course, is characterized by an equivalence class under
diffeomorphisms of solutions of the equations of motion; but we do not
know how to effectively coordinatize the space of the equivalence
classes in terms of realistically observable quantities.  The
individual observables that we use in concrete applications are just a
small number, and are far from capturing the full gauge invariant
physics.  The difficulty of writing a complete class of observables is
in fact well known (see \cite{obsGR,obs,io,Earman,don}, and references
therein), and raises several well known problems.  For instance it
complicates the canonical analysis of the theory and it is a serious
obstacle to quantization.  Furthermore, it generates conceptual
difficulties for the very physical interpretation of the theory
\cite{Earman}.

Attempts to define a complete class of observables abound in the
literature.  It is easier to construct such observables in the
presence of matter that in the context of pure GR. This is because the
difficulty of writing observables is the consequence of the absence of
absolute localization in a general relativistic theory.  If there is
matter we can localize things with respect to the matter.  For
instance, we can consider GR interacting with four scalar matter
fields.  Assume that the configuration of the fields is sufficiently
nondegenerate.  Then the components of metric field at points defined
by given values of the matter fields are observable.  The idea is
clearly the same as defining an observable as the distance between us
and a planet: instead of having just a few planets to define a few
physical distances, we imagine to have a continuous of matter so that
physical distances are defined everywhere.  This idea has been
developed in a number of variants, such as dust carrying clocks and
others, by many authors, including the present one.  See
\cite{obs,io,don} and references therein.  One succeeds in
constructing a complete set of observables, but the extent to which
the result is realistic or useful is certainly questionable.  It is
rather unsatisfactory to understand the theory in terms of fields that
do not exist, or phenomenological objects such as dust, and it is
questionable whether these procedures could make sense in the quantum
theory, where the aim is to describe Planck scale.  Other (earlier)
attempts to write a complete set of observables are in the context of
pure GR \cite{obsGR}.  The idea is to construct four scalar functions
of the metric (say, scalar polynomials of the curvature), and use
these to localize points.  The value of a fifth scalar function in a
point where the four scalar functions have a given value is an
observable.  This works, but the result is mathematically very
intricate and physically extremely unrealistic.  It is certainly
possible, in principle, to construct detectors of such observables,
but I doubt any experimenter would get founded for proposing to build
such apparata.  Do we thus have to declare defeat and, in a general
relativistic context, give up the hope of having a complete set of
well defined observables, which are realistic, easy to construct, and
do not assume that the world is different from what it is?

In this paper, I'd like to argue that we do not have to declare
defeat.  I propose a simple way out, based on GR coupled with a
minimal and {\em very\/} realistic amount of additional matter. 
Indeed, the solution I discuss is so realistic that it is in fact
real: it was inspired by an already existing technology, the Global
Positioning System (GPS), the first technological application of GR,
or the first large scale technology that needs to take GR effects into
account \cite{GPS}.  The other source of the ideas presented here is
the Null Surface Formulation (NSF) approach to GR \cite{NSF}.  What
follows can be seen as a sort of pedestrian version of some NSF ideas. 
Ideas related to the ones presented here have been presented by
Massimo Pauri and Luca Lusanna \cite{luca}.  See also \cite{Bahder}.

The idea here is simple.  Consider a general covariant system formed
eucaby GR coupled with four small bodies.  These are taken to have
negligible mass; they will be considered as point particles for
simplicity, and called ``satellites".  Assume that the four satellites
follow timelike geodesics; that these geodesics meet in a common
(starting) point $O$; and, that at $O$ they have a given (fixed) speed
--the same for the four-- and directions as the four vertices of a
tetrahedron.  The theory might include any other matter.  Then (there
is a region $\cal R$ of spacetime for which) we can uniquely associate
four numbers $s^\alpha, \alpha=1,2,3,4$ to each spacetime point $p$ as
follows.  Consider the past lightcone of $p$.  This will (generically)
intersect the four geodesics in four points $p_{\alpha}$.  The numbers
$s^\alpha$ are defined as the distance between $p_{\alpha}$ and $O$. 
We can use the $s^\alpha$'s as physically defined coordinates.  The
components $g_{\alpha\beta}(s)$ of the metric tensor in these
coordinates are observable quantities.  They are invariant under
four-dimensional diffeomorphisms (because, of course, these deform
the metric as well as the satellites' worldlines).  They define a
complete set of observables for the region $\cal R$.

The physical picture is simple, and its realism is transparent. 
Imagine that the four ``satellites" are in fact satellites, each
carrying a clock that measures the proper time along its trajectory,
starting at the meeting point $O$.  Imagine also that each satellite
broadcasts its local time with a radio signal.  Suppose I am at the
point $p$ and have an electronic device that simply receives the four
signals and displays the four readings.  See Figure 1.  These four
numbers are precisely the four physical coordinates $s^\alpha$ defined
above.  Current technology permits to perform these measurements with
accuracy well within the relativistic regime \cite{GPS,appli}.  If I
then use a rod, and a clock and measure the physical distances between
$s^\alpha$ coordinates points, I am directly measuring the components
of the metric tensor in the physical coordinate system.  In the
terminology of Ref.\,\cite{partial}, the $s^\alpha$'s are {\em
partial\/} observables, while $g_{\alpha\beta}(s)$ are {\em
complete\/} observables.

As shown below, the physical coordinates $s^\alpha$ have nice
geometrical properties; they are characterized by
\begin{equation}
    g^{\alpha\alpha}(s)=0, \ \ \ \ \ \alpha=1,\ldots,4. 
    \label{eq:gauge}
\end{equation}
Surprisingly, in spite of the fact that they are defined by what looks
like a rather nonlocal procedure, the evolution equations for
$g_{\alpha\beta}(s)$ are local.  These evolution equations can be
written explicitly using the Arnowitt-Deser-Misner (ADM) variables
\cite{ADM}.  Lapse and Shift turn out to be fixed local functions of
the three metric.

\begin{figure}
\centerline{{\psfig{figure=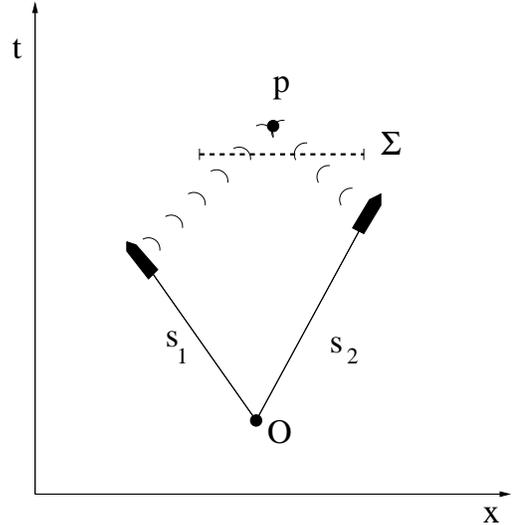,height=7cm}}}
\bigskip \caption{$s_{1}$ and $s_{2}$ are the GPS coordinates of the
point $p$.  $\Sigma$ is a Cauchy surface with $p$ in its future domain
of dependence.}
\end{figure}

In what follow, we begin for simplicity by introducing the GPS
coordinates $s^\alpha$ in Minkowski space.  This allows us to
introduce some tools in a simple context.  Then we go over to a
general spacetime.  We take the speed of light to be one, signature
$[+,-,-,-]$, and we assume the Einstein summation convention only for
couples of repeated indices that are one up and one down.  Thus
$\alpha$ is not summed over in (\ref{eq:gauge}).  While dealing with
Minkowski spacetime, the spacetime indices $\mu,\nu$ are raised and
lowered with the Minkowski metric.  We write an arrow over three- as
well as four-dimensional vectors.

\vskip.5cm

Consider a tetrahedron in three-dimensional euclidean space.  Let its
center be at the origin and it four vertices $\vec v^{\alpha}$, where
and $\vec v^{\alpha} \cdot \vec v^{\beta}=-1/3$ for $\alpha\ne\beta$,
have unit length $|\vec v^{\alpha}|^2=1$.  Here $\alpha=1,2,3,4$ is an
index that distinguishes the four vertices, and should not be confused with
vector indices.  With a convenient orientation, these vertices have
polar coordinates $u=(r,\theta,\phi)$
\begin{eqnarray}
    v^{1u} & = & (1,0,0) \\
    v^{2u} & = & (1,\alpha,0) \\
    v^{3u} & = & (1,\alpha,2\pi/3) \\
    v^{4u} & = & (1,\alpha,-2\pi/3)
    \label{eq:polar}
\end{eqnarray}
with $\cos\alpha=-1/3$, and cartesian coordinates ($a=1,2,3$) 
\begin{eqnarray}
    v^{1a} & = & (0,\ 0,\  1) \\
    v^{2a} & = & ({2\sqrt{2}}/{3},\  0,\ -{1}/{3}) \\
    v^{3a} & = & (-{\sqrt{2}}/{3},\  \sqrt{2/ 3},\ -{1}/{3}) \\
    v^{4a} & = & (-{\sqrt{2}}/{3},\ -\sqrt{2/3},\ -{1}/{3}).
    \label{eq:cartesian}
\end{eqnarray}
Let us now go to a four dimensional Minkowski space.  Consider four
timelike 4-vectors $\vec W^\alpha$, of length one, $|\vec
W^{\alpha}|^2=1$, representing the
normalized 4-velocities of four particles moving away from the origin
in the directions $\vec v^\alpha$ at a common speed $v$.  
Their Minkowski coordinates $(\mu=0,1,2,3)$ are 
\begin{equation}
    W^{\alpha\mu} = \frac{1}{\sqrt{1-v^2}}\, (1,\ v\,  
    v^{\alpha a}). 
\end{equation}
We fix the velocity $v$ by requiring the determinant of the matrix
$W^{\alpha\mu}$ to be one one.  (This choice fixes $v$ at about half
the speed of light; a different choice changes only a few
normalization factors in what follows.)  The four by four matrix
$W^{\alpha\mu}$ plays an important role in what follows.  Notice that
it is a fixed matrix whose entries are certain given numbers.

Consider one of the four 4-vectors, say $\vec W=\vec W^{1}$.  Consider
a free particle in Minkowski space that starts from the origin with
4-velocity $\vec W$.  Call it a ``satellite".  Its world line $l$ is
$\vec x(s)=s\vec W$.  Since $\vec W$ is normalized, $s$ is precisely
the proper time along the world line.  Consider now an arbitrary point
$p$ in Minkowski spacetime, with coordinates $\vec X$.  We want to
compute the value of $s$ at the intersection between $l$ and the past
light cone of $p$.  The calculation is particularly easy in a Lorentz
frame in which the satellite stays still, namely has vanishing
3-velocity, and $p$ is in the $(t,x)$ plane, with coordinates $(T,X)$. 
Assume $X>0$. In this frame, the equation of $l$ is
\begin{equation}
    x=0, 
\end{equation}
the proper time is $s=t$, and the equation of the (intersection with
the $(t,x)$ plane of the) past light cone of $p$ is
\begin{equation}
    (X-x)=\pm(T-t),\ \ \ \ \ \ \ t<T.
\end{equation}
Taking all this together gives 
\begin{equation}
    s =T-X. 
\end{equation}
Notice now that in this frame $|\vec X|^2=T^2-X^2$ and $\vec X\cdot \vec 
W=T$, so that we can write
\begin{eqnarray}
T-X &=& T - \sqrt{T^2-(T^2-X^2)} \nonumber \\
&=& \vec X\cdot \vec W - \sqrt{(\vec X\cdot
\vec W)^2-|\vec X|^2}.
\end{eqnarray}
From the last two equations we obtain 
\begin{equation}
s = \vec X\cdot \vec W - \sqrt{(\vec X\cdot \vec W)^2-|\vec X|^2}.
\end{equation}
But this equation is Lorentz covariant and therefore it is true in any
Lorentz system.  (A direct full four dimensional calculation in an 
arbitrary system gives the same result.)

Let us now consider four satellites, moving out of the origin at
4-velocity $\vec W^{\alpha}$.  If they radio broadcast their position,
an observer at the point $p$ with Minkowski coordinates $\vec X$ 
receives the four signals $s^\alpha$
\begin{equation}
s^\alpha = \vec X\cdot \vec W^\alpha - \sqrt{(\vec X\cdot \vec
W^\alpha)^2-|\vec X|^2}.
\label{coordinates}
\end{equation}  
We introduce (non-Lorentzian) general coordinates $s^\alpha$ on
Minkowski space, defined by the change of variables
(\ref{coordinates}).  These are the coordinates read out by a GPS
device in Minkowski space.  The Jacobian matrix of the change of
coordinates is given
by 
\begin{equation}
\frac{\partial s^\alpha}{\partial x^\mu} = 
W_{\mu}^\alpha - \frac{ W_{\mu}^\alpha(\vec X\cdot \vec
W^\alpha)-X_{\mu}}{\sqrt{(\vec X\cdot \vec W^\alpha)^2-|\vec X|^2}}, 
\end{equation}  
where $W_{\mu}^\alpha$ and $X_{\mu}$ are $W^{\alpha\mu}$ and $X^\mu$ 
with the spacetime index lowered with the Minkowski metric.  This
defines the the (co-)tetrad field $E_{\mu}^\alpha(s)$
\begin{equation}
E_{\mu}^\alpha(s(X)) = \frac{\partial s^\alpha}{\partial x^\mu}(X).
\end{equation}  
The contravariant metric tensor is given by $g^{\alpha\beta}(s)=
E_{\mu}^\alpha(s) E^{\mu\beta}(s)$.  Using $|\vec W^{\alpha}|^2=1$, 
a straightforward calculation shows that
\begin{equation} 
    g^{\alpha\alpha}(s)=0, \ \ \ \ \ \alpha=1,\ldots,4. 
    \label{gauge}
\end{equation}
This equation has the following nice geometrical interpretation.  Fix
$\alpha$ and consider the one-form field $\omega^\alpha=ds^\alpha$. 
In $s^\alpha$ coordinates, this one-form has components
$\omega^\alpha_{\beta} = \delta^\alpha_{\beta}$, and therefore
``length" $|\omega^\alpha|^2= g^{\beta\gamma}\omega^\alpha_{\beta}
\omega^\alpha_{\gamma} = g^{\alpha\alpha}$.  But the ``length" of a 1-form
is proportional to the volume of the (infinitesimal, now) 3-surface
defined by the form.  The 3-surface defined by $ds^\alpha$ is the
surface $s^{\alpha}=constant$.  But $s^{\alpha}=constant$ is the set
of points that read the GPS coordinate $s^\alpha$, namely that
receive a radio broadcasting from a same event $p_{\alpha}$ of the
satellite $\alpha$, namely that are on the future light cone of
$p_{\alpha}$.  Therefore $s^{\alpha}=constant$ is a portion of this
light cone, therefore it is a null surface, therefore its volume is
zero, therefore $|\omega^\alpha|^2=0$, therefore $
g^{\alpha\alpha}=0$.

Since the $s^\alpha$ coordinates define $s^{\alpha}=constant$ surfaces
that are null, we denote them as ``null GPS coordinates".  It is
useful to introduce another set of GPS coordinates as well, which have
the traditional timelike and spacelike character.  We denote these as
$s^\mu$, call them ``timelike GPS coordinates", and define them by
\begin{equation}
 s^\alpha = W_\mu^{\alpha} s^\mu. 
    \label{eq:timelike}
\end{equation}  
This is a simple algebraic relabeling of the names of the four GPS
coordinates, such that $s^{\mu=0}$ is timelike and $s^{\mu=a}$ is
spacelike.  In these coordinates, the gauge condition (\ref{gauge})
reads
\begin{equation}
   W_\mu^{\alpha} W_\nu^{\alpha} g^{\mu\nu}(s)=0. 
    \label{eq:timelikegauge}
\end{equation} 
This can be interpreted geometrically as follows.  The (timelike) GPS
coordinates are coordinates $s^\mu$ such that the four 1-forms fields
\begin{equation}
    \omega^\alpha=W^\alpha_{\mu}ds^\mu
    \label{omega}
\end{equation}
are null.  

Let us now jump from Minkowski space to full GR. Consider GR coupled
with four satellites of negligible mass that move geodesically and
whose world lines emerge from a point $O$ with directions and velocity
as above.  Locally around $O$ the metric can be taken to be
Minkowskian; therefore the details of the initial conditions of the
satellites worldlines can be taken as above.  The phase space of this
system is the one of pure GR plus 10 parameters, giving the location
of $O$ and the Lorentz orientation of the initial tetrahedron of
velocities.  The integration of the satellites' geodesics and of the
light cones can be arbitrarily complicated in an arbitrary metric. 
However, if the metric is sufficiently regular, there will still be a
region $\cal R$ in which the radio signals broadcasted by the
satellites are received.  (In case of multiple reception, the
strongest one can be selected.  That is, if the past light cone of $p$
intersects $l$ more than once, generically there will be one
intersection which is at shorter luminosity distance.)  Thus, we still
have well defined physical coordinates $s^\alpha$ on $\cal R$. 
Equation (\ref{gauge}) holds in these coordinates, because it depends
just on the properties of the light propagation around $p$.  We define
also timelike GPS coordinates $s^\mu $ by (\ref{eq:timelike}), and we
get the condition (\ref{eq:timelikegauge}) on the metric tensor.

To study the evolution of the metric tensor in GPS coordinates it is
easier to shift to ADM variables $N,N^a,\gamma_{ab}$.  These are
functions of the covariant components of the metric tensor, defined in
general by
\begin{eqnarray}
    ds^2 & = & g_{\mu\nu}dx^\mu dx^\nu \nonumber \\ 
   & = & N^2 dt^2 - \gamma_{ab}(dx^a-N^a dt)(dx^b-N^b dt). 
\end{eqnarray}
Equivalently, they are related to the contravariant components of the 
metric tensor by 
\begin{equation}
    g^{\mu\nu}v_{\mu}v_{\nu}=-\gamma^{ab}v_{a}v_{b}+(n^\mu v_{\mu})^2,
    \label{eq:contro}
\end{equation}
where $\gamma^{ab}$ is the inverse of $\gamma_{ab}$ and 
$n^\mu=(1/N,N^a/N)$. 
Using these variables, the gauge condition (\ref{eq:timelikegauge}) 
reads 
\begin{equation}
   W_a^{\alpha} W_b^{\alpha} \gamma^{ab}=(W^\alpha_\mu n^\mu)^2. 
\end{equation}
Notice now that his can be solved for the Lapse and Shift as a
function of the three-metric (recall that $W^\alpha_\mu $ are fixed 
numbers), obtaining
\begin{equation}
   n^\mu = W^\mu_\alpha q^\alpha
\end{equation} 
where $W^\mu_\alpha$ is the inverse of the matrix $W_\mu^\alpha$ and
\begin{equation}
q^\alpha = \sqrt{W_a^{\alpha} W_b^{\alpha} \gamma^{ab}}. 
\label{q}
\end{equation}
Or, explicitly, 
\begin{equation}
    N  = \frac{1}{W^0_\alpha q^\alpha}, \hspace{.4cm}
    N^a =  \frac{W^a_\alpha q^\alpha}{W^0_\alpha q^\alpha}. 
    \label{eq:Na}  
\end{equation}
The geometrical interpretation is as follows.  We want the 1-form
$\omega^\alpha$ defined in (\ref{omega}) to be null, namely its norm
to vanish.  But in the ADM formalism this norm is the sum of two
parts: the norm of the pull back of $\omega^\alpha$ on the constant
time ADM surface, which is $q^\alpha$, given in (\ref{q}), and depends
on the three metric; plus the square of the projection of
$\omega^\alpha$ on $n^\mu$.  We can thus obtain the vanishing of the
norm by adjusting the Lapse and Shift.  We have four conditions (one
per each $\alpha$) and we can thus determine Lapse and Shift out of
three metric.  In other words, whatever is the three metric, we can
always adjust Lapse and Shift so that the gauge condition
(\ref{eq:timelikegauge}) is satisfied. 

But in the ADM formalism, the arbitrariness of the evolution in the
Einstein equations is entirely captured by the freedom in choosing
Lapse and Shift.  Since here Lapse and Shift are uniquely determined
by the three metric, evolution is determined uniquely if the initial
data on a Cauchy surface are known.  Therefore the evolution in the
GPS coordinate $s^0$ of the GPS components of the metric tensor,
$g_{\mu\nu}(s)$, is governed by deterministic equations: the ADM
evolution equation with Lapse and Shift determined by equations
(\ref{q}--\ref{eq:Na}).  Notice also that evolution is local, since
the ADM evolution equations, as well as the equations
(\ref{q}--\ref{eq:Na}), are local.\footnote{This does not imply that
the full set of equations satisfied by $g_{\mu\nu}(s)$ must local,
since initial conditions on $s^0=0$ satisfy four other constraints
besides the ADM ones.}

To understand what is going on in this gauge, consider an arbitrary
coordinate system $x^\mu$, and an arbitrary coordinate transformation
$s^\alpha = s^\alpha(x)$ such that the transformed metric tensor
satisfies (\ref{gauge}).  This is equivalent to imposing one
differential equation on each of the four functions $s^\alpha(x)$, as
follows
\begin{equation}
    g^{\mu\nu}(x)\partial_{\mu}s^\alpha(x)\partial_{\nu}s^\alpha(x)=0.
    \label{diffeq}
\end{equation}
Equivalently, the problem is to find four independent exact null
1-forms $\omega^\alpha=ds^\alpha$, integrable over a finite region. 
Or, equivalently again, the problem is to find four nowhere-parallel
foliations of Minkowski space with families of null surfaces.  The
last is precisely the main ingredient of the Null Surface Formulation
of GR \cite{NSF}.  A 10 parameter family of solutions of this problem
is given by the construction above, namely by the light fronts
emerging from the geodesics of the satellites, and labelled with the
proper time along these geodesics.  Equation (\ref{diffeq}) has of
course more solutions than these, since there are more families of
null foliations than the ones constructed here.  A null surface in a
solution of (\ref{diffeq}), for instance, does not need to focus in a
point.  Even less, the surfaces of the family need to focus on a line
which is a geodesic, and so on.  But if we require that the surfaces
focus on a point, that the points are along a timelike geodesic, that
the label of the foliation is given by the proper time along this
geodesic, and that the four geodesics meet at a single point and with
the prescribed angles, then the space of solutions of (\ref{diffeq})
is reduces down to a {\em finite\/} 10 parameter space.  The 10
parameters are the location of the origin $O$ and the Lorentz
orientation of the tetrahedron of initial velocities.  (By pushing $O$
all the way to past infinity, one obtains the asymptotic NSF
coordinates.)

How can the evolution of the quantities $g_{\mu\nu}(s)$ be local?  The
conditions on the null surfaces described in the previous paragraph
are nonlocal.  Coordinate distances yield typically to nonlocality:
Imagine we define physical coordinates in the solar system using the
cosmological time $t_{c}$ and the spatial distances
$x_{S},x_{E},x_{J}$ (at fixed $t_{c}$) from, say, the Sun the Earth
and Jupiter.  The metric tensor $g_{\mu\nu}(t_{c},x_{S},x_{E},x_{J})$
in these coordinates is observable, but its evolution is highly non
local.  To see this, imagine that in this moment (in cosmological
time), Jupiter is swept away by a huge comet.  Then the value of
$g_{\mu\nu}(t_{c},x_{S}, x_{E},x_{J})$ here changes instantaneously,
without any local cause: the value of the coordinate $x_{J}$ has
changed because of an event happened far away.  What's special about
the GPS coordinates that avoids this nonlocality?  The answer is that
the value of a GPS coordinate in a point $p$ does in fact depend on
what happens ``far away" as well.  Indeed, it depends on what happens
to the satellite.  However, it only depends on what happened to the
satellite when it was broadcasting the signal received in $p$, and
this is in the past of $p$ !  If $p$ is in the past domain of
dependence of a partial Cauchy surface $\Sigma$, then the value of
$g_{\mu\nu}(s)$ in $p$ is completely determined by the metric an its
derivative on $\Sigma$, namely evolution is causal, because the entire
information needed to set up the GPS coordinates is in the data in
$\Sigma$.  See Figure 1.  Explicitly, the $s^\alpha=constant$ surfaces
around $\Sigma$ can be uniquely integrated ahead all the way to $p$. 
They certainly can, as they represent just the evolution of a light
front!  This is how local evolution is achieved by these coordinates.

\vskip.5cm

Let us summarize.  We have introduced a set of physical coordinates,
determined by certain material bodies.  Geometrical quantities such as
the components of the metric tensor expressed in physical coordinates
are of course observable.  This procedure is well known.  The novelty
here is that we have shown that in order to obtain physical
coordinates, there is no need, as usually assumed, to introduce a
large unrealistic amount of matter or to construct complicated and
unrealistic physical quantities out of the metric tensor.  Indeed,
four particles are sufficient to coordinatize a (region of a)
four-geometry.  Furthermore, the coordinatization procedure is not
artificial: it is the real one utilized by existing technology.

Notice that the degrees of freedom of the theory we have considered
are still only two per point (plus the 10 parameters of the initial
condition of the satellites.)  This is why initial data on $s^0$ for
$g_{\mu\nu}(s)$ must satisfy additional constraints.  We leave the
study of these and the explicit construction of the associated
Hamiltonian formalism for future investigations.

We conclude with some simple comments on observability.  The
components of the metric tensor in (timelike) GPS coordinates can be
measured in principle as follows.  Take a rod of physical length $L$
(small with respect to the distance along which the gravitational
field changes significatively) with two GPS devices at its ends
(reading timelike GPS coordinates).  Orient the rod (or search among
recorded readings) so that the two GPS devices have the same reading
$s$ of all coordinates except for $s^1$.  Let $\delta s^1$ be the
difference in the two $s^1$ readings.  Then we have along the rod
\begin{equation}
    ds^2=g_{11}(s)\delta s^1 \delta s^1=L^2. 
\end{equation}
Therefore 
\begin{equation}
g_{11}(s)=(L/\delta s^1)^2.
\end{equation}
Non-diagonal components of $g_{ab}(s)$ can be measured by simple
generalizations of this procedure. The $g_{0b}(s)$ are then
algebraically determined by the gauge conditions.  In a thought
experiment, a spaceship could travel in a spacetime region and compose
a map of values of the GPS components of the metric tensor.  

To avoid confusion, it is may be useful to recall here the distinction
between partial and complete observables \cite{partial}.  A partial
observable is a quantity to which a measuring procedure can be
associated.  A complete observable is an observable quantity that can
predicted by the theory, or, equivalently, whose knowledge provides
information on the state of the system.  For instance, in the theory
of a single harmonic oscillator $q(t)$ the time $t$ and the position
of the oscillator $q$ are partial observables, while the position of
the oscillator at a given time, or at the initial time, is a complete
observable.  The GPS coordinates are partial observables: we can
associate them a measuring procedure (this is what has been done in
this paper), but we can of course not ``predict" $s^1$.  Equivalently,
there is no sense in which the metric field of the universe is
characterized by a certain value of $s^1$.  The complete observables,
or true observables, are the quantities $g_{\mu\nu}(s)$, for any given
value of the coordinates $s^\mu$.  These quantities are diffeomorphism
invariant, are uniquely determined by the initial data and in a
canonical formulation are represented by functions on the phase space
that commute with all constraints.

Finally, notice that the observables we have defined are a
straightforward generalization of Einstein's ``point coincidences"%
\footnote{``All our space-time verifications invariably amount to a
determination of space-time coincidences.  If, for example, events
consisted merely in the motion of material points, then ultimately
nothing would be observable but the meeting of two or more of these
points.  \cite{Him}}.  In a sense, they are {\em precisely\/}
Einstein's point coincidences.  Einstein's ``material points" are just
replaced by photons (light pulses): the spacetime point $s^\alpha$ is
characterized as the meeting point of four photons designated by the
fact of carrying the radio signals $s^\alpha$.

\vskip.5cm

I thank Don Marolf for a crucial exchange, Serge Lazzarini for useful
comments on an early draft of this paper, John Earman and John Norton
for a long term enlightening discussion on observability in GR,
Margherita Fiani for explaining to me the technical details on the way
the GPS actual functions and Enzo Marinari for motivating me to
develop this work.  This work was partially supported by NSF Grants
PHY-9900791.

\end{document}